\documentclass[10pt, conference]{IEEEtran}
\IEEEoverridecommandlockouts
\usepackage{cite}
\usepackage{amsmath,amssymb,amsfonts}
\usepackage{algorithmic}
\usepackage{graphicx}
\usepackage{textcomp}
\usepackage{xcolor}
\def\BibTeX{{\rm B\kern-.05em{\sc i\kern-.025em b}\kern-.08em
    T\kern-.1667em\lower.7ex\hbox{E}\kern-.125emX}}

\usepackage{algorithm}
\usepackage{algorithmic}
\usepackage{soul}

\usepackage{multirow}
\usepackage{multicol}
\usepackage{array}

\usepackage{pifont}

\usepackage[normalem]{ulem}

\usepackage{amsmath}    
\usepackage{courier}
\usepackage{makecell}
\usepackage{graphicx}    
\usepackage[table]{xcolor}
\usepackage{booktabs}     
\usepackage{colortbl}   
\usepackage{caption}  
\usepackage[caption=false]{subfig}
\usepackage{balance}
\usepackage[colorlinks=true, urlcolor=black, linkcolor=black, citecolor=black, pdfborder={0 0 0}]{hyperref}

\begin{document}

\title{CASS-RTL: Correctness-Aware Subspace Steering for RTL Generation with LLMs}

\author{\IEEEauthorblockN{Mohammad Akyash, Nowfel Mashnoor, Kimia Azar, Hadi Kamali}
\IEEEauthorblockA{\textit{Department of Electrical and Computer Engineering (ECE), University of Central Florida, Orlando, FL 32816, USA} \\
\{mohammad.akyash, nowfel.mashnoor, azar, kamali\}@ucf.edu}
}

\maketitle

\begin{abstract}

Recent advances in large language models (LLMs) have enabled the automatic synthesis (generation) of register-transfer level (RTL) code from natural language instructions, offering a promising pathway to accelerate chip design. Unlike typical natural language (and software coding) tasks, LLM-based RTL code generation demands strict cycle accuracy with concurrency, where minor logical errors can render a circuit unusable or insecure. While prior work has explored hallucination mitigation via external verification, self-evaluation prompts, retrieval-augmented prompting, domain specific fine-tuning, agentic solutions, and reasoning, these approaches largely overlook the attention-oriented internal mechanisms of LLMs that may inherently correlate with RTL correctness. This work proposes CASS-RTL, a first-of-its-kind framework for discovering and leveraging LLMs' correctness-aware components to guide RTL generation toward functionally accurate outputs. We (i) identify attention heads whose activation patterns consistently differentiate correct from incorrect RTL; (ii) construct a low-dimensional subspace capturing correctness-relevant signals; and (iii) design a lightweight, geometry-aware intervention that steers the model at inference time. CASS-RTL is fully model-agnostic, requires no additional supervision or retraining, and readily integrates into existing models. Empirically, we evaluate CASS-RTL on multiple models and observe 10\%--20\% improvement in pass@1/5/10 accuracy on VerilogEval and 5\% improvement on CVDP, demonstrating the effectiveness of our method in enhancing reliability without sacrificing model efficiency or requiring a large labeled dataset for fine-tuning\footnote{Code is available at \url{https://github.com/mhakyash/CASS-RTL}}.

\end{abstract}


\section{Introduction}

Large language models (LLMs) have recently shown remarkable promise in generating register-transfer level (RTL) hardware descriptions from natural language instructions, unlocking new capabilities for intelligent hardware design and rapid semiconductor development \cite{liu2024rtlcoder, cui2024origen, akyash2025rtl, liu2024craftrtl, zhao2024codev, zhao2024mage, deng2025scalertl, zhu2025codev, akyash2025simeval, Khan2025sagehls, mashnoor2025llm, mashnoor2026language}. However, unlike natural language generation, RTL code synthesis demands strict adherence to cycle-accurate functional correctness with concurrency, structural constraints, and semantic precision \cite{akyash2025decortl}. While LLMs can produce syntactically valid RTL, their outputs sometimes contains subtle functional bugs. In some cases, the generated RTL, though structurally and syntactically correct, fails to align with the given instruction commonly known as \textit{hallucination} \cite{ping2025hdlcore}, \cite{yang2025haven}. This misalignment poses a serious barrier to deploying LLMs in hardware design, where even minor semantic deviations (from the instructions) can lead to invalid or insecure implementations \cite{akyash2024selfhwdebug, rezakhani2026safetune}. 


\begin{figure*}[t]
  \centering
  \includegraphics[width=\linewidth]{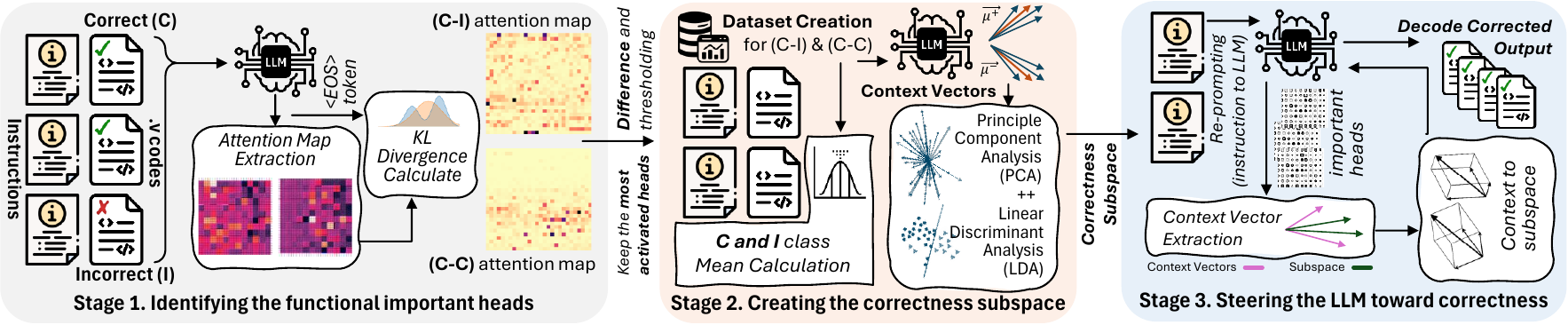}
  \vspace{-10pt}
  \caption{Overall Framework of the CASS-RTL}
  \label{fig:framework}
  \vspace{-10pt}
\end{figure*}

While Li et al. \cite{li2023halueval} shows that incorporating external knowledge (e.g. search engine–based retrieval \cite{shuster2022language} or knowledge graphs \cite{wu2023retrieve}) can help reduce hallucinations\footnote{Follow-ups like HaluEval-Wild show that even very strong models (e.g., GPT-4 Turbo) still hallucinate on $\sim$20\% of prompts \cite{zhu2024halueval}.} in LLM outputs, growing evidence from natural language tasks suggests that \textbf{\ul{LLMs possess internal signals indicating when their outputs are hallucinated or low-quality}}, with few recent studies exploring and demonstrating that \textit{“\ul{models often know when they are wrong\footnote{A trained classifier on hidden states of GPT-J shows 71–83\% accuracy at predicting whether an LLM response is true vs. false \cite{azaria2023internal}.}},”} even if their responses appear fluent and confident \cite{li2023iti}, \cite{zhang2024truthx}, \cite{meng2022locating}, \cite{azaria2023internal}. Leveraging this, recent works show that LLMs can self-evaluate to detect hallucinations without human supervision. For instance, Zhang et al. \cite{zhang2024self} prompt models to assess the factuality of their own responses (self-refine-oriented), improving reliability through self-alignment and confidence calibration. Other works go further by using self-generated critiques as feedback to iteratively refine model outputs, enabling LLMs to improve through self-feedback \cite{madaan2023self}, \cite{ji2023towards}. These key findings indicate that LLMs embed truth-relevant information in their internal representations, but this latent knowledge is often underutilized during standard inference. To better harness this latent capability, a promising line of research focuses on guiding LLMs through their internal representations. For instance, TruthX \cite{zhang2024truthx} explores disentangling truthfulness from semantics in the latent space using contrastive learning, while contrast-consistent search (CCS) \cite{burns2022discovering} identifies internal directions correlated with truthful behavior without requiring labeled data. Other methods, such as inference-time intervention (ITI) \cite{li2023iti} analyze intermediate activations (e.g. ranking attention heads) to localize and influence truth-relevant components within the model. 

In chip design process (especially at RTL coding) LLMs are becoming an increasingly prominent \cite{liu2024rtlcoder}. However, current methods overwhelmingly emphasize surface-level improvements, through fine-tuning \cite{cui2024origen}, agentic prompting \cite{zhao2024mage}, and self-verification \cite{huang2024towards}, while largely overlooking the internal signals models already possess about correctness. Consequently, LLM-based RTL generation inherits the same fundamental limitation seen in natural-language tasks that \textit{“\ul{models often know when they generate wrong RTL}”}. Prior methods offer insights in natural-language settings but do not directly transfer to RTL code, which lacks semantic contrast and demands fine-grained control. As a result, RTL generation remains black-box and difficult to steer, motivating the need to leverage model-internal reasoning signals for correct RTL synthesis.

To address this gap, we propose \textbf{CASS-RTL}, \textit{Correctness-Aware Subspace Steering for RTL generation}, a framework that identifies correctness-relevant latent directions within the model and applies lightweight, geometry-aware interventions at inference time (without additional training) to improve the functional correctness of LLM-generated RTL code. Our main contributions are as follows:

\begin{table}[b]

\centering
\scriptsize
\vspace{-5pt}
\setlength{\tabcolsep}{1pt}
\caption{Steering Methods for Enhancing LLM Correctness.}
\label{tab:steering_methods_comparison}
\begin{tabular*}{\linewidth}{@{\extracolsep{\fill}} lcccc }
\toprule
\textbf{Method} & \textbf{Training Req.} & \textbf{Label Dependency} & \textbf{Compt. Cost} & \textbf{Binary Decisioning} \\
\cmidrule(r){1-1}\cmidrule(lr){2-2}\cmidrule(lr){3-3}\cmidrule(lr){4-4}\cmidrule(l){5-5}
ITI \cite{li2023iti}                & Medium & Truth              & Medium & No \\
TruthX \cite{zhang2024truthx}       & High   & Truth / Semantics  & High   & No \\
Truth Forest \cite{chen2024truthforest} & Medium & Truth          & Medium & No \\
CCS \cite{burns2022discovering}     & No     & None               & Low    & Yes \\
\cmidrule(r){1-1}\cmidrule(lr){2-2}\cmidrule(lr){3-3}\cmidrule(lr){4-4}\cmidrule(l){5-5}
\textbf{CASS-RTL}                   & \textbf{No}     & \textbf{None}               & \textbf{Low}    & \textbf{No}  \\
\bottomrule
\end{tabular*}
\end{table}

\noindent \ul{\textit{\textbf{(i) Correctness-relevant signal exploit.}}} We identify attention heads whose intermediate activations correlate with correct RTL, revealing internal structures that encode correctness.

\noindent \ul{\textit{\textbf{(ii) Correctness-aware latent subspace.}}} We construct a geometry-aligned subspace from intermediate activations that captures model-internal signals correlated with correctness.

\noindent \ul{\textit{\textbf{(iii) Inference-time steering mechanism.}}} We introduce a geometry-aware steering method that adjusts model activations along the learned subspace, enabling fine-grained control over generation.

\noindent \ul{\textit{\textbf{(iv) Extensive empirical validation.}}} On the VerilogEval \cite{liu2023verilogeval} and CVDP \cite{pinckney2025cvdp} benchmarks, CASS-RTL significantly improves functional correctness across diverse RTL tasks, outperforming strong baselines.

\section{Related Work}

\begin{figure*}
\centering
\includegraphics[width=\linewidth]{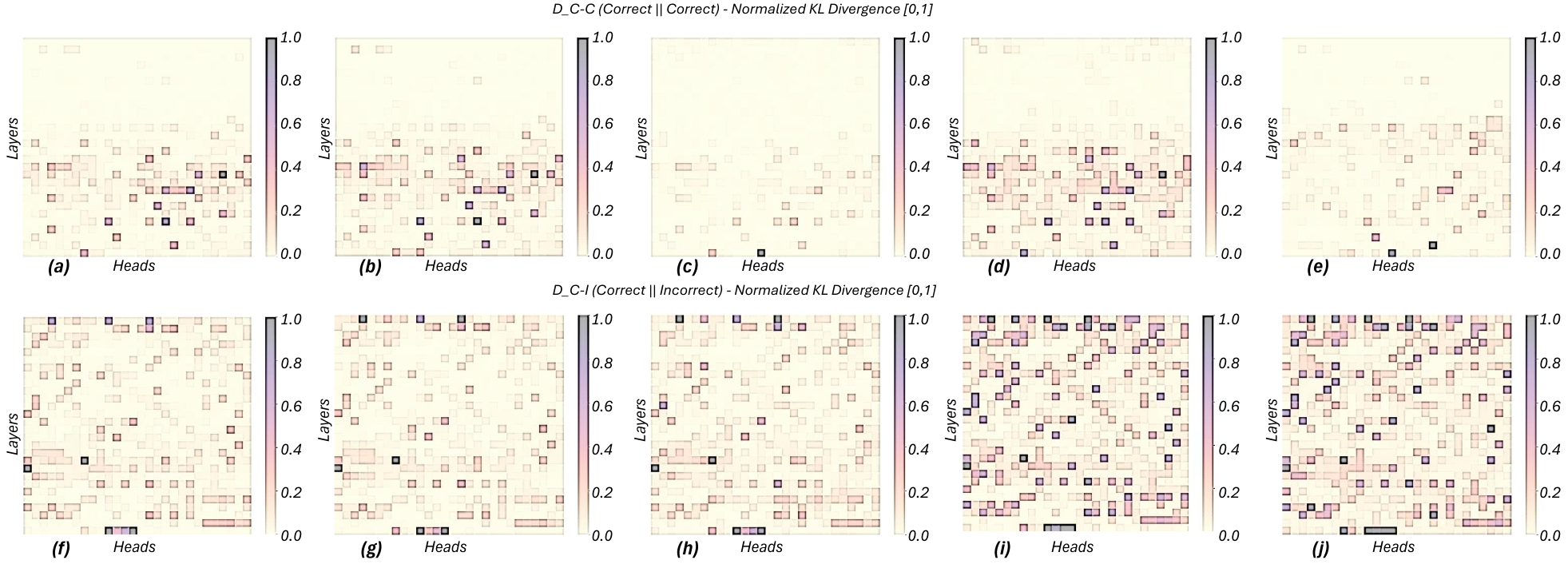}
\vspace{-20pt}
\caption{KL Divergence Heatmaps for 5 Example Pairs. (a)–(e) show divergence between functionally correct implementations (C–C), while (f)–(j) show divergence between correct and incorrect implementations (C–I). Each heatmap represents layer–head divergence values, with \textbf{darker shades indicating higher divergence}. C–I maps show broader and more pronounced divergence validates the hypothesis that functional correctness influences attention more than stylistic variation.}
\vspace{-15pt}
\label{fig:divergence_heatmaps}
\end{figure*}

Several works investigate leveraging internal LLM representations to improve truthfulness and reduce hallucinations \cite{meng2022locating, mashnoor2026meltrtl}. These methods largely target natural language tasks and aim to isolate or manipulate latent components associated with correctness \cite{li2023iti, zhang2024truthx, burns2022discovering, chen2024truthforest}.

TruthX \cite{zhang2024truthx} disentangles truthfulness from semantics using a contrastive autoencoder, but requires curated truth-labeled pairs and may overlook fine-grained RTL semantics. To avoid labeled data, CCS \cite{burns2022discovering} identifies truth-related directions through contrastive consistency, although its formulation is tailored to binary linguistic tasks and does not naturally extend to structured RTL generation. ITI \cite{li2023iti} instead ranks attention heads using linear probes and performs inference-time intervention, but still depends on labeled data and assumes linear separability. Building on ITI, Truth Forest (TrFr) \cite{chen2024truthforest} employs multi-dimensional probes to capture richer intermediate features, yet it likewise relies on truth-labeled data and assumptions that may not generalize to domain-specific outputs such as RTL.

Table~\ref{tab:steering_methods_comparison} summarizes key differences. Our method identifies correctness sensitive heads in a structure-aware manner, requires no truth/false annotations, and is designed specifically for structured RTL generation where semantics and correctness are tightly coupled. By aggregating divergence signals across diverse functional tasks and constructing a correctness-aware subspace, we enable fine-grained, interpretable, geometry-driven steering suitable for complex hardware design domains.

\section{CASS-RTL: Methodology}
\label{sec:method}

Rather than treating RTL generation as a purely black-box decoding task, CASS-RTL, as shown in Figure \ref{fig:framework}  analyzes how the model internally behaves when producing correct versus incorrect RTL implementations. This section shows our methodology for identifying correctness sensitive attention heads in decoder-only LLMs and leveraging their representations to construct a subspace for correctness-aware steering at inference. 

\subsection{Attention Mechanism}

We consider decoder-only transformers with multi-head self-attention (MHSA) at each layer. Given an input of \(S\) tokens with embeddings in \(\mathbb{R}^d\), the hidden states are \(X \in \mathbb{R}^{S \times d}\). At layer \(l\) and head \(h\), attention is computed as:
\begin{equation} \label{eq:attention}
Q_{l,h}=XW_{l,h}^Q,\quad K_{l,h}=XW_{l,h}^K,\quad V_{l,h}=XW_{l,h}^V,
\end{equation}
with \(W_{l,h}^{Q,K,V}\in\mathbb{R}^{d\times d_h}\). Scaled dot-product attention and head outputs are:
\begin{equation} \label{eq:softmax}
A_{l,h}=\text{softmax}\!\left(\frac{Q_{l,h}K_{l,h}^\top}{\sqrt{d_h}}\right)\!\in\mathbb{R}^{S\times S},
\end{equation}
\begin{equation} \label{eq:head_output}
O_{l,h}=A_{l,h}V_{l,h}\in\mathbb{R}^{S\times d_h}.
\end{equation}
MHSA concatenates head outputs and applies a projection:
\begin{equation} \label{eq:mhsa}
\text{MHSA}_l(X)=\text{Concat}(O_{l,1},\dots,O_{l,H})W_l^O.
\end{equation}

We focus on the final token (\(\langle\text{EOS}\rangle\)) at position \(j=S-1\). For each head \((l,h)\), its contextual embedding is:
\begin{equation} \label{eq:embed}
\mathbf{o}_{l,h}:=o_{l,h,S-1}\in\mathbb{R}^{d_h}.
\end{equation}
As the final token attends to all previous tokens, \(\mathbf{o}_{l,h}\) provides a compact summary of the sequence-level state and is used for correctness analysis.
\begin{figure}[b]
    \centering    
    \vspace{-10pt}
    \includegraphics[width=\linewidth]{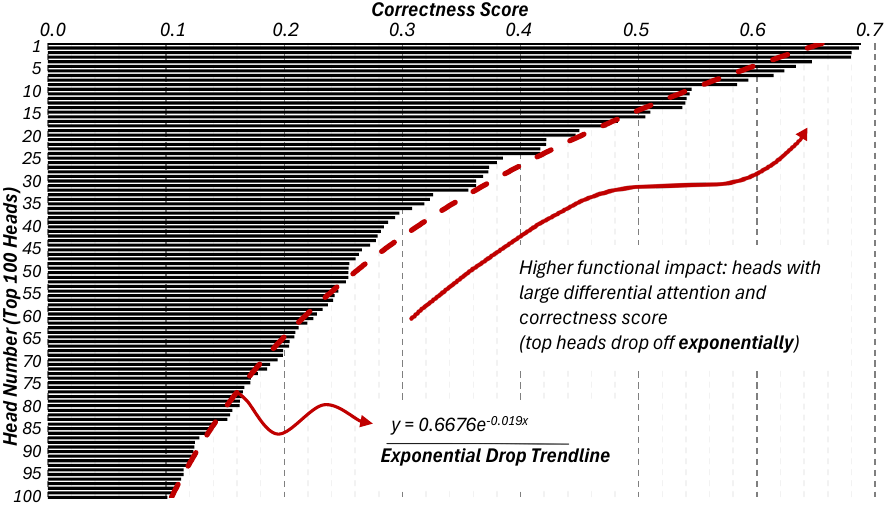}
    \caption{Top-100 Head Correctness Score}
    \label{fig:head_scores}
\end{figure}

\subsection{Identifying Key Heads for Correctness}

Our methodology in CASS-RTL begins by identifying specific attention heads in decoder-only, autoregressive LLMs (the main target model is Qwen2.5-Coder-14B) that are especially sensitive to functional correctness in RTL code generation. These models generate tokens left-to-right, with each token conditioned on the entire preceding sequence. We hypothesize that attention heads behave differently when processing functionally correct versus incorrect RTL code, even if both are syntactically valid. To capture the model's final decision behavior, we extract the attention vectors directed toward the $\langle \text{EOS} \rangle$, which encapsulate how each head integrates information across the input when producing the last token. For a given generation $G$ with sequence length $S_G$, the attention vector from layer $l$ and head $h$ is denoted $\mathbf{v}_{l,h}^{(G)} \in \mathbb{R}^{S_G}$. To investigate functional sensitivity, we curate paired RTL examples across various basic building blocks\footnote{~The basic building blocks of RTL are relatively limited and well-structured, and all RTL designs are ultimately composed of hierarchical combinations of these primitives.} (e.g., control elements, memory structs, combinatorial and arithmetic modules), categorized as follows:

\noindent \ul{\textit{\textbf{(i) Correct vs. Incorrect (C-I).}}} A functionally correct ($C$) and an incorrect ($I$) design for the same instruction.

\noindent \ul{\textit{\textbf{(ii) Correct vs. Correct (C-C).}}} Two structurally different but functionally equivalent implementations ($C_A$, $C_B$).

We treat each attention vector as a probability distribution over input tokens and compute the Kullback–Leibler (KL) divergence between them as demonstrated in Eq. \ref{kl_div}:
\begin{equation} \label{kl_div}
\begin{aligned}
    D_{\text{C-I}}^{(T_i)}(l,h) &= D_{\text{KL}}(P_{l,h}^{(C)} \parallel P_{l,h}^{(I)}) \\
    D_{\text{C-C}}^{(T_i)}(l,h) &= D_{\text{KL}}(P_{l,h}^{(C_A)} \parallel P_{l,h}^{(C_B)})
\end{aligned}
\end{equation}
Although both divergences reflect differences in attention behavior, the C-I divergence captures the model’s sensitivity to semantic/functional errors, whereas C-C divergence highlights variations due to stylistic/formatting differences that leave functionality unchanged. To isolate functional sensitivity from superficial variation, we compute the difference:
\begin{equation}
    S_{\text{correctness}}^{(T_i)}(l,h) = D_{\text{C-I}}^{(T_i)}(l,h) - D_{\text{C-C}}^{(T_i)}(l,h)
\end{equation}
This subtraction removes background noise caused by stylistic diversity which demonstrate heads that react \textit{specifically} to correctness violations rather than just any difference (isolating variation that is more strongly linked to functional semantics). A high value of $S_{\text{correctness}}^{(T_i)}(l,h)$ means that head $(l,h)$ distinguishes functional errors more sharply than it does stylistic variation (suggesting that it encodes reliability-critical cues such as dependency mismatches or illegal signal interactions) which makes it a strong candidate for correctness-aware analysis. We aggregate these correctness-oriented differences across $M$ samples by averaging:
\begin{equation}
    \overline{S}_{\text{correctness}}(l,h) = \frac{1}{M} \sum_{i=1}^{M} S_{\text{correctness}}^{(T_i)}(l,h)
\end{equation}

As shown in Figure~\ref{fig:divergence_heatmaps}, C-I divergence maps often exhibit stronger and more spatially distributed activation patterns compared to C-C maps. This supports our hypothesis that certain heads attend more distinctly to functional correctness than to surface-level code differences. To quantify this sensitivity, we compute a correctness score for each attention head based on its divergence behavior. Figure~\ref{fig:head_scores} presents the top-100 correctness scores ($S_\text{correctness}$), demonstrating a long-tailed distribution where only a subset of heads exhibit strong correctness-aware behavior. We then apply a head selection strategy: top-$k$ ranking, where we retain the $k$ heads with the highest positive scores. These selected heads form our steering subspace in the inference-time correction pipeline.

\subsection{Correctness-Aware Subspace Steering}

Building on the identification of correctness-sensitive attention heads, we propose a subspace-based intervention method that steers the LLM during inference toward functionally correct RTL code generation. While prior works such as ITI~\cite{li2023iti} and TruthX~\cite{zhang2024truthx} demonstrate direction-based interventions for general-purpose tasks, our approach introduces a class-conditional, multi-head subspace specifically designed for the structured semantics of RTL. Let $\mathcal{H} = \{(l_1,h_1), (l_2,h_2), \ldots, (l_{N_{\text{sel}}}, h_{N_{\text{sel}}})\}$ denote the set of $N_{\text{sel}}$ previously identified correctness-aware heads. For each head $(l,h) \in \mathcal{H}$ and generation $G$, we extract the value (context) vector at the end-of-sequence $\langle \text{EOS} \rangle$ token:
\begin{equation}
\mathbf{o}_{l,h,\text{eos}}^{(G)} \in \mathbb{R}^{D_h}.
\end{equation}

These are concatenated into a unified representation:

\begin{equation}
\mathbf{X}^{(G)} = \left[ \mathbf{o}_{l_1, h_1, \text{eos}}^{(G)} \, \Big| \, \dots \, \Big| \, \mathbf{o}_{l_{N_{\text{sel}}}, h_{N_{\text{sel}}}, \text{eos}}^{(G)} \right] \in \mathbb{R}^{D_{\text{total}}},
\end{equation}

where $D_{\text{total}} = N_{\text{sel}} \cdot D_h$. We organize the extracted representations into two datasets: $\mathcal{D}^+ = \{ \mathbf{X}^{(G)} \mid G \text{ is correct} \}$ and $\mathcal{D}^- = \{ \mathbf{X}^{(G)} \mid G \text{ is incorrect} \}$. The class means are computed as demonstrated in Eq. \ref{eq:class_means}. We then center each dataset by subtracting its class mean, and pool the centered vectors for subspace construction. To extract a low-dimensional correctness-aware subspace, we apply principal component analysis (PCA) to the pooled matrix. PCA identifies orthogonal directions that capture the dominant modes of variation across both correct and incorrect activations, which are directions that often correspond to latent factors the model uses to encode structural or semantic distinctions. Formally, we solve Eq. \ref{eq:pca_formula}, where $\Sigma$ is the empirical covariance matrix, and $B \in \mathbb{R}^{D_{\text{total}} \times k}$ contains the top-$k$ eigenvectors. 
\begin{equation}
\label{eq:class_means}
\mu^+ = \frac{1}{|\mathcal{D}^+|} \sum_{\mathbf{X} \in \mathcal{D}^+} \mathbf{X}, \quad \mu^- = \frac{1}{|\mathcal{D}^-|} \sum_{\mathbf{X} \in \mathcal{D}^-} \mathbf{X}.
\end{equation}
\begin{equation}
\label{eq:pca_formula}
B = \arg\max_{B^\top B = I} \operatorname{Tr}(B^\top \Sigma B),
\end{equation}

\begin{figure}[t]
  \centering
  \includegraphics[width=\linewidth]{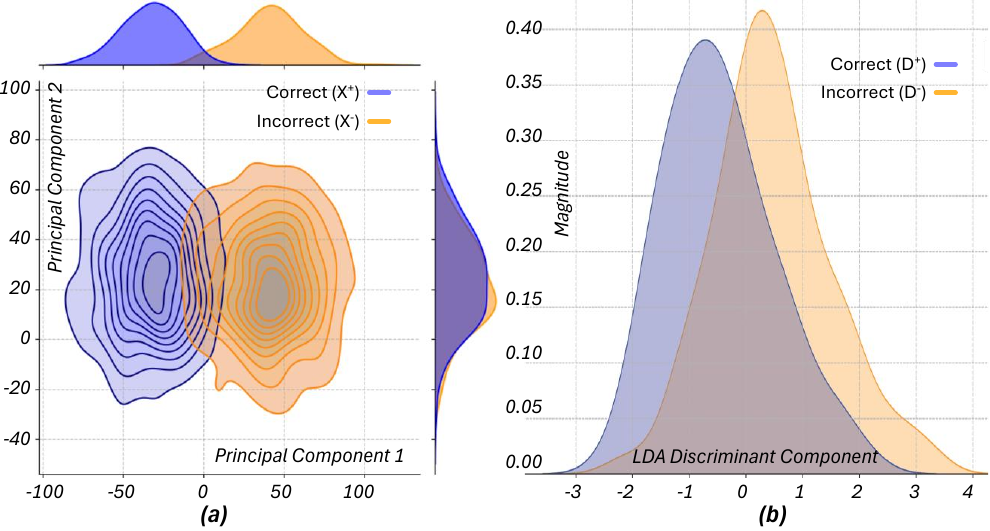}
  \vspace{-15pt}
  \caption{(a) PCA projection of multi-head representations. Correct and incorrect samples form separable clusters, showing that functional correctness is embedded in the latent structure; (b) LDA projection of correctness-conditioned head representations. The distributions for correct and incorrect generations show strong separation, indicating that correctness is linearly decodable from the subspace.}
  \label{fig:pca_lda_projection}
  \vspace{-15pt}
\end{figure}

This PCA-based subspace filters out noise and captures discriminative structural patterns relevant to correctness. We visualize the distribution of samples in the top-2 PCA components in Figure~\ref{fig:pca_lda_projection} (a). Even in this unsupervised projection, correct and incorrect generations form visibly separable clusters, highlighting that structural correctness is reflected in the latent space. To further validate separability, we apply linear discriminant analysis (LDA) to project $\mathbf{X}^{(G)}$ into a 1D space that maximizes class separation. As shown in Figure~\ref{fig:pca_lda_projection} (b), correct ($\mathcal{D}^+$) and incorrect ($\mathcal{D}^-$) samples are linearly separable with minimal overlap. This reinforces the effectiveness of direction-based interventions.

In terms of inference-time steering, at each decoding step during autoregressive generation, we extract the current multi-head activation vector $\mathbf{X}_{\text{current}} \in \mathbb{R}^{D_{\text{total}}}$ and project it onto the correctness subspace, shown in Eq. \ref{eq:x_proj}. Then, as illustrated in Eq. \ref{eq:diff_x}, we compute a correction (differential-based) vector.
\begin{equation}\label{eq:x_proj}
\mathbf{X}_{\text{proj}} = B B^\top \mathbf{X}_{\text{current}}, \quad
\mu_U^+ = B B^\top \mu^+.
\end{equation}
\begin{equation}\label{eq:diff_x}
\Delta \mathbf{X} = \mu_U^+ - \mathbf{X}_{\text{proj}},
\end{equation}
This differential-based correction vector is followed by updating the representation as shown in Eq. \ref{eq:x_steered}, where $\alpha > 0$ is a tunable steering strength. The updated vector is then reshaped and injected back into the same attention heads at their respective layers.
\begin{equation}\label{eq:x_steered}
\mathbf{X}_{\text{steered}} = \mathbf{X}_{\text{current}} + \alpha \cdot \Delta \mathbf{X},
\end{equation}

Unlike ITI, which globally shifts the residual stream, or TruthX, which builds static per-token projection layers, our method preserves per-head semantics and aligns with the model’s architecture. It supports fine-grained, token-wise interventions that are especially suited for RTL’s compositional logic. As RTL code correctness is structurally constrained, this makes the internal representations highly reflective of functional outcomes. Our subspace method leverages these structural regularities to guide generation without modifying model weights. By nudging activations toward the correct-class mean, we reliably improve output validity, offering a principled mechanism for correctness-aware decoding.

\section{Experiments}

We evaluate the effectiveness of correctness-aware steering for RTL generation, focusing on functional correctness via simulation-based verification. We compare CASS-RTL against baseline decoding and state-of-the-art steering methods across multiple model families, including  Codellama 7B, QwenCoder-2.5 14B and the fine-tuned CodeV model \cite{zhao2024codev}, enabling analysis on both pretrained and domain-adapted models and highlighting orthogonality to approaches such as fine-tuning. 
All experiments are conducted on VerilogEval \cite{liu2023verilogeval} and CVDP \cite{pinckney2025cvdp}. We also include causal and representational analyses to better understand the effectiveness of correctness-aware steering and its impact on model behavior.

\begin{figure}[b]
    \centering
    \vspace{-10pt}
    \includegraphics[width=\linewidth]{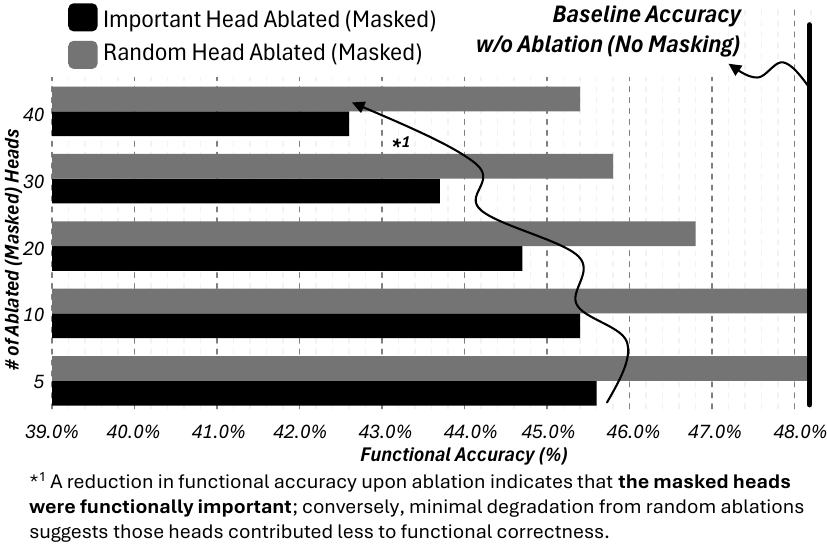}
    \caption{Effect of ablating different sets of heads on functional correctness. Ablating important heads causes a larger drop in RTL accuracy compared to random ablation.}
    \label{fig:head_ablation}
\end{figure}

\subsection{Comparison to the State-of-the-art}

To evaluate our steering strategy, we reimplemented ITI~\cite{li2023iti}, a standard activation-intervention method. ITI locates attention heads whose internal activations correlate with functionally correct outputs and intervenes on those heads during decoding. Our pipeline follows two steps. First, we extracted activation vectors from every attention head at the final token position and labeled them based on whether the generated RTL sample was functionally correct. Binary classifiers were trained per head, and we selected the top $k=30$ heads with the highest correctness-predictive accuracy. For each selected head, we computed a direction vector by subtracting the mean activation of incorrect samples from that of correct ones. Second, at inference, these directions were injected to nudge the corresponding head activations toward the ``correct'' side without modifying model weights or requiring retraining.

Table~\ref{tab:method_comparison_func} reports functional correctness across models and top-$k$ decoding on VerilogEval and CVDP. CASS-RTL ($\alpha=0.3$) consistently outperforms both base models and ITI across all settings. While ITI provides moderate gains via iterative decoding optimization, it remains less effective than our subspace-driven approach, which directly targets correctness-sensitive directions in internal representations. Improvements are larger on stronger models, suggesting that richer correctness-oriented structure is better exploited. CASS-RTL requires no auxiliary classifiers, is geometry-aware, and generalizes across model sizes and decoding settings, yielding a more robust and scalable alternative to direction-based interventions.

\begin{table}[t]
\centering
\scriptsize
\setlength{\tabcolsep}{3pt}
\caption{RTL functional correctness (Func@\emph{k}) across models.}
\label{tab:method_comparison_func}
\vspace{-7pt}
\begin{tabular*}{\linewidth}{@{\extracolsep{\fill}} l lccc ccc ccc }
\toprule
\multirow{2}{*}{\textbf{Dataset}} & \multirow{2}{*}{\textbf{Method}} 
& \multicolumn{3}{c}{\textbf{CodeLlama 7B}} 
& \multicolumn{3}{c}{\textbf{QwenCoder 14B}}
& \multicolumn{3}{c}{\textbf{CodeV \cite{zhao2024codev}}} \\
& 
& \multicolumn{3}{c}{\textbf{(Pre-trained)}} 
& \multicolumn{3}{c}{\textbf{(Pre-trained)}}
& \multicolumn{3}{c}{\textbf{(RTL Fine-tuned)}} \\
\cmidrule(lr){3-5}
\cmidrule(lr){6-8}
\cmidrule(lr){9-11}
& 
& @1 & @5 & @10 
& @1 & @5 & @10 
& @1 & @5 & @10 \\
\midrule

\multirow{3}{*}{\textbf{VerilogEval}}
& Base 
& 18.2 & 22.7 & 24.3 
& 37.1 & 44.8 & 50.6 
& 53.2 & 65.1 & 68.5 \\

& ITI  
& 24.3 & 26.9 & 27.5 
& 44.8 & 46.7 & 51.2 
& 57.6 & 67.3 & 70.0 \\

& \textbf{CASS-RTL} 
& \textbf{28.8} & \textbf{30.1} & \textbf{31.4}
& \textbf{48.7} & \textbf{52.5} & \textbf{56.4}
& \textbf{62.8} & \textbf{70.0} & \textbf{73.7} \\

\midrule

\multirow{3}{*}{\textbf{CVDP}}
& Base 
& 3.8 & 3.8 & 3.8 
& 10.2 & 11.5 & 14.1 
& 5.1 & 6.4 & 6.4  \\

& ITI  
& 3.8 & 3.8 & 5.1 
& 12.8 & 14.1 & 14.1 
& 7.6 & 8.9 & 8.9 \\

& \textbf{CASS-RTL} 
& \textbf{5.1} & \textbf{6.4} & \textbf{6.4}
& \textbf{16.6} & \textbf{19.2} & \textbf{20.5} 
& \textbf{8.9} & \textbf{8.9} & \textbf{10.2} \\

\bottomrule
\end{tabular*}
\vspace{-10pt}
\end{table}


\begin{figure}[t]
\centering
\includegraphics[width=\linewidth]{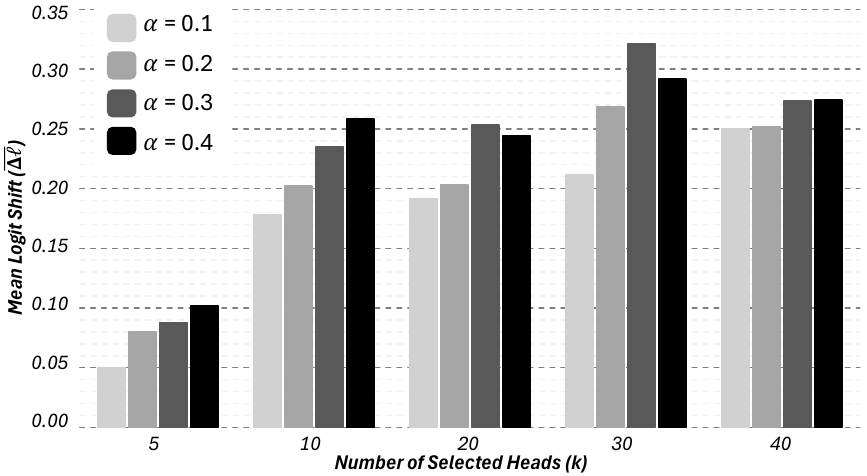}
\caption{Effect of steering strength \( \alpha \) on mean logit shift \( \overline{\Delta \ell} \). Increasing \( \alpha \) consistently boosts model confidence in tokens.}
\vspace{-10pt}
\label{fig:bar-logit-shift}
\end{figure}

\begin{figure}[b]
\centering
\vspace{-15pt}
\includegraphics[width=\linewidth]{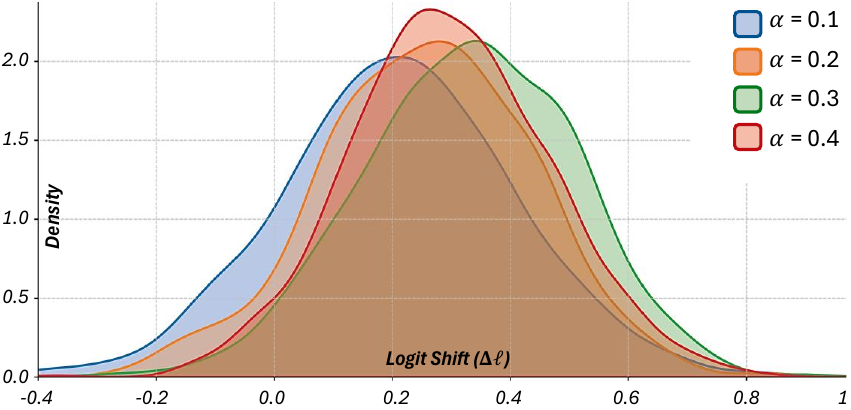}
\caption{Distribution of token-level logit shifts \( \Delta \ell \) across varying steering strengths \( \alpha \). Higher values of \( \alpha \) shift the distribution toward more positive logit gains.}
\label{fig:dist-logit-shift}
\end{figure}

\begin{figure}[h]
    \centering
    \includegraphics[width=\linewidth]{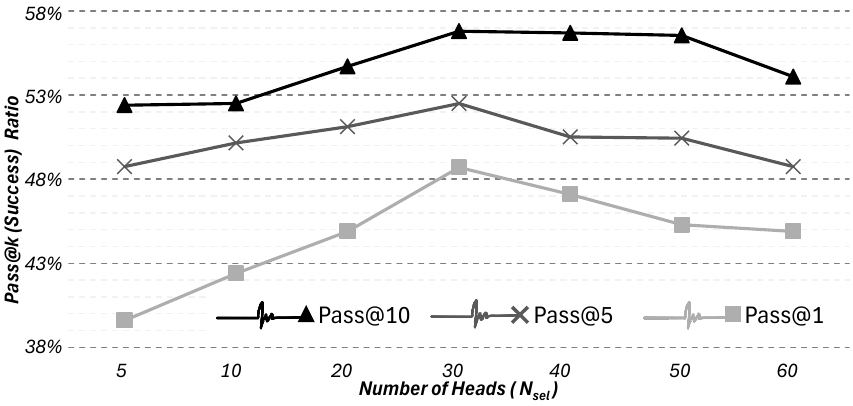}
    \vspace{-20pt}
    \caption{Effect of varying the number of selected heads \( N_{\text{sel}} \) on Pass@k accuracy using the QwenCoder 2.5 14B model on VerilogEval. Accuracy improves up to \( N_{\text{sel}} = 30 \), after which performance degrades slightly, indicating over-steering.}
    \vspace{-10pt}
    \label{fig:nsel_ablation}
\end{figure}

\subsection{Causal Intervention via Targeted Head Mask}

Although our divergence-based metric correlates selected heads with RTL correctness, correlation alone does not imply causation. To test whether these heads are \emph{causally} important, we perform an ablation study in which their outputs are masked during inference and the resulting impact on RTL generation is measured. We evaluate three ablation conditions to assess the causal role of correctness-sensitive heads:

\noindent \ul{\textit{\textbf{(i) Baseline.}}} No attention heads are masked. This setting serves as the reference point, reflecting the model’s native generation.

\noindent \ul{\textit{\textbf{(ii) Ablation of Important Heads.}}} All \(N_{\text{sel}}\) heads with the highest \(\overline{S}_{\text{correctness}}\) scores are masked at inference. This tests whether the heads identified by our scoring method are \emph{functionally necessary} for correctness; if these heads truly encode correctness-sensitive structure, disabling them should significantly degrade performance.

\noindent \ul{\textit{\textbf{(ii) Ablation of Random Heads.}}} \(N_{\text{sel}}\) attention heads are randomly selected and masked. This is for the general effect of removing attention capacity, ensuring that any performance drop cannot be attributed merely to reduced model expressivity. Functional correctness is evaluated using the same simulation-based verification pipeline on a fixed RTL test set. As shown in Figure~\ref{fig:head_ablation}, masking important heads significantly degrades functional accuracy relative to the baseline, indicating a failure to maintain logical consistency and signal behavior. In contrast, masking an equal number of random heads causes only a small drop, suggesting that reduced attention capacity alone does not explain the degradation. This gap demonstrates that the identified heads are causally involved in correct RTL generation, validating our selection strategy based on $\overline{S}_{\text{correctness}}$.

\subsection{Steering Effectiveness: Logit-Based View}

To evaluate the impact of our correctness-aware steering mechanism on the model’s predictive behavior, we examine how steering modifies the model’s confidence in emitting the ground-truth RTL tokens throughout the decoding process. Confidence is reflected directly in the token-level logits, which determine the probability distribution over the vocabulary at each step. For each validation instruction paired with its reference RTL implementation, we therefore generate two decoding trajectories: one using the unmodified model and one using CASS-RTL steering. At each decoding step \( i \), we record the logit assigned to the ground-truth token \( t_i \) in both settings. Let \( \ell_{\text{orig}}(t_i) \) and \( \ell_{\text{steered}}(t_i) \) denote the logits before and after steering, respectively. We define the logit shift as:
\begin{equation}
\Delta \ell(t_i) = \ell_{\text{steered}}(t_i) - \ell_{\text{orig}}(t_i)
\end{equation}
A positive value of \( \Delta \ell(t_i) \) indicates that steering increased the model’s confidence in the correct token. To summarize the overall effect, we compute the mean logit shift over the sequence via Eq. \ref{eq:delta_logit}, where \( T \) is the total number of ground-truth tokens.
\begin{equation}\label{eq:delta_logit}
\overline{\Delta \ell} = \frac{1}{T} \sum_{i=1}^{T} \Delta \ell(t_i)
\end{equation}

Figure~\ref{fig:bar-logit-shift} shows how the average logit shift increases with stronger steering parameters \( \alpha \), supporting our hypothesis that steering improves alignment with correct generations. To analyze steering effects at a finer granularity, Figure~\ref{fig:dist-logit-shift} illustrates the distribution of token-level logit shifts for different \( \alpha \) values. As \( \alpha \) increases, the distributions shift rightward, indicating that a greater proportion of tokens benefit from steering. This analysis further validates the efficacy of our subspace-based intervention in enhancing functional correctness during RTL code generation.

\subsection{Impact of Heads on Pass@k Accuracy}

We analyze CASS-RTL sensitivity to the number of correctness-sensitive heads used for steering. All experiments here use QwenCoder 2.5 14B on VerilogEval, varying the number of selected heads \(N_{\text{sel}} \in \{5, 10, 20, 30, 40\}\) and evaluating Pass@1, Pass@5, and Pass@10. Results are shown in Figure~\ref{fig:nsel_ablation}. Pass@1 and Pass@5 improve steadily as \(N_{\text{sel}}\) increases, peaking around \(N_{\text{sel}} = 30\). Beyond this point, accuracy drops slightly, indicating that adding weaker heads introduces noise into the steering signal. Pass@10 continues to rise and saturates at \(N_{\text{sel}} \approx 35\text{--}40\), reflecting greater robustness under wider decoding beams. Overall, the results indicate a clear \emph{sweet spot} in the number of heads used for intervention and show that overly large selections can dilute correctness information.

\section{Acknowledgment}

This work was supported in part by AMD University Program (Graduate Research Fellowship).

\section{Conclusion}

We proposed CASS-RTL, a correctness-aware subspace steering framework that leverages internal LLM representations to improve RTL generation. By identifying correctness-sensitive heads and constructing a low-dimensional subspace, CASS-RTL enables lightweight, training-free decoding interventions. Across models and benchmarks (VerilogEval, CVDP), it achieves substantial functional gains with negligible overhead. These improvements are orthogonal to model scaling, dataset specialization, and fine-tuning. Beyond correctness, CASS-RTL also improves syntactic validity, increases confidence on ground-truth tokens, and exhibits strong causal alignment.

\balance
\bibliographystyle{IEEEtran}
\bibliography{references}

\end{document}